# COMMENTARY ON INDIA'S ECONOMY AND SOCIETY SERIES

# 11

## WORLD DEVELOPMENT REPORT 2020: TRADING FOR DEVELOPMENT IN THE AGE OF GLOBAL VALUE CHAINS

Rajkumar Byahut
Sourish Dutta
Chidambaran G. Iyer
Manikantha Nataraj

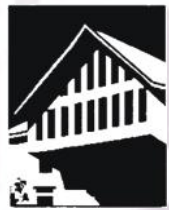

CDS
Thiruvananthapuram

India's Economy and indeed its society has been undergoing a major change since the onset of economic reforms in 1991. Overall growth rate of the economy has increased, the economy is getting increasingly integrated with the rest of the world and public policies are now becoming very specific compared over arching framework policies of the pre-reform period. Over the past few years, a number of important policies have been enunciated, like for instance the policy on moving towards a cashless economy to evolving a common market in the country through the introduction of a Goods and Services Tax. Issues are becoming complex and the empirical basis difficult to decipher. For instance the use of payroll data to understand growth in employment, origin-destination passenger data from railways to understand internal migration, Goods and Services Tax Network data to understand interstate trade. Further, new technologies such as Artificial Intelligence, Robotics and Block Chain are likely to change how manufacturing and services are going to be organised. The series under the "Commentary on India's Economy and Society" is expected to demystify the debates that are currently taking place in the country so that it contributes to an informed conversation on these topics. The topics for discussion are chosen by individual members of the faculty, but they are all on issues that are current but continuing in nature. The pieces are well researched, engages itself sufficiently with the literature on the issue discussed and has been publicly presented in the form of a seminar at the Centre. In this way, the series complements our "Working Paper Series".

CDS welcomes comments on the papers in the series, and these may be directed to the individual authors.

# WORLD DEVELOPMENT REPORT 2020: TRADING FOR DEVELOPMENT IN THE AGE OF GLOBAL VALUE CHAINS

**Rajkumar Byahut**
**Sourish  Dutta**
**Chidambaran   G. Iyer**
**Manikantha Nataraj**

# World Development Report 2020:
# Trading for Development in the Age of Global Value Chains

**1. Background**

    The importance of trade to an economy needs no emphasis. You sell products or services that you are competitive at and buy those where you are not. Experience of countries such as South Korea and China demonstrate that resources required for development can be garnered through trade; thus, motivating many countries to embrace trade as a means for development. Simultaneously, emergence of 'Global Value Chain' or 'GVC' as they are popularly known has changed the way we trade. Though the concept of GVC was introduced in the early 2000s, there are examples of global value chains before the 1980s. However, the scale of the phenomenon and the way in which technological change – by lowering trade costs – has allowed fragmentation of production was not possible before (Hernandez et al, 2014). In this context, the World Bank has recently published its 'World Development Report 2020: Trading for Development in the Age of Global Value Chains' (WDR). The report prescribes that GVCs still offer developing countries a clear path to progress and that developing countries can achieve better outcomes by pursuing market-oriented reforms specific to their stage of development. The report mentions that participation in GVCs can deliver a double dividend. First, firms are more likely to specialize in the tasks in which they are most productive. Second, firms are able to gain from connections with foreign firms, which pass on the best managerial and technological practices. As a result, countries enjoy faster income growth and falling poverty (WDR, 2020). Given the current context of the Indian economy – slowing of growth, increasing labour force, and need for generating for more employment – implementing the recommendations of the WDR probably seems like the need of the hour. However, as it often happens in public policy, there is no one size that fits all. This commentary will attempt to critically evaluate the WDR so as to distil out lessons for the Indian context. We focus on four issues from the WDR,

    a.    Measurement and trend of GVCs participation

    b.    Drivers of participation in GVCs

    c.    Transition or Upgrading in GVCs

    d.    Consequences for Development

    The layout of the article is as follows, in the next section we first introduce the concept of GVC, followed by the issue of measurement of GVCs, trends in GVC participation are discussed next. The third section lays out the drivers of participation. The issue of transition or upgrading in GVCs is

picked up next followed by the section on the consequences for development. These sections not only summarize the above mentioned issues from the report but also try to bring in the Indian context. This is then followed by observations and conclusion.

## 2. Measurement and Trend of GVCs

### 2.1 Measurement of GVCs

A value chain can be defined simply as the full range of activities that firms and workers do to bring a product from its conception to its end use and beyond. Typically, a value chain includes the following activities: design, production, marketing, distribution and support to the final consumer. These activities can be performed within the same firm or divided among different firms. Global value chains are thus value chains that are increasingly spread over several countries (Hernandez et al, 2014). As per WDR (2020), GVC is a series of stages in the production of a product or service, for sale to consumers. Each stage adds value, and at least two stages are in different countries. For example, bicycle manufacturing is heavily reliant on trade. They are assembled using parts and components from all over the world, especially Asia and Europe (Figure 1). Bianchi, a bicycle manufacturer, carries out all of its design, prototyping, and conception work in Italy, and then assembles most of its bicycles in Taiwan and China, using parts and components from China, Italy, Japan, Malaysia, and many other parts of the world. Each parts producer has niche expertise, which cutting across product lines is true for most of the GVCs. Assembling a bicycle from parts and components made around the world improves efficiency and results in a cheaper and higher-quality bicycle for the consumer. As a result of this extensive bicycle value chain, the trade in bicycle parts in recent years has outstripped the trade in bicycles by 15-25 percent (WDR, 2020).

**Figure 1: Value chain in Bicycles**

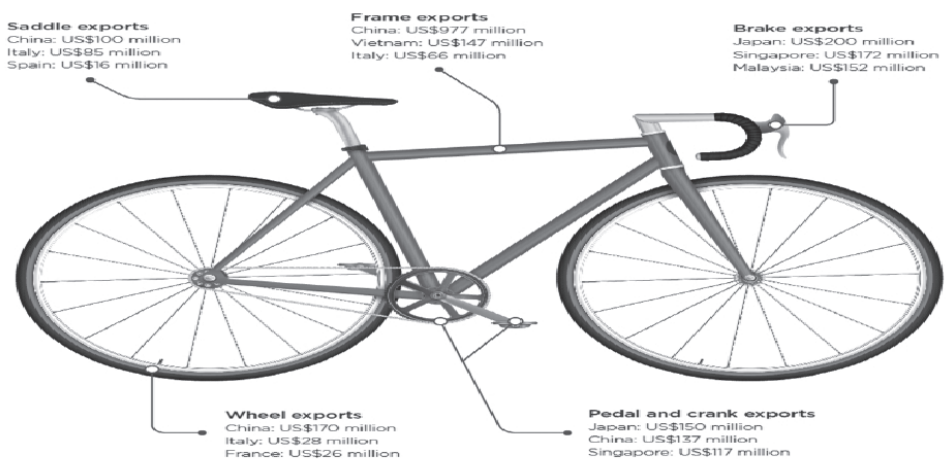

Source: WDR, 2020

The spread of GVCs across the world has led to efforts at understanding GVCs. One of the steps in this direction has been to measure GVCs and to position them in the globalised production process. Unlike traditional trade, goods and services in GVC trade cross borders multiple times, generally, requiring more than two countries. Moreover, trade is more in the form of intermediate inputs rather than final goods. Traditional trade statistics such as customs data that are generally used to measure standard trade is inadequate to measure GVC trade. The concept of 'import to export' (Baldwin & Lopez-Gonzalez 2015), i.e. use of imported content (in terms of value-added embodied in materials, intermediate inputs or tasks) in the exports at the country or industry or firm level is one such measure that can be used to measure GVC participation at the country or industry or firm level.

Data requirements to measure the GVC phenomenon are intense. For example, global input-output tables (involving customs data and national input-output tables) have now been generated to measure GVCs. These global or inter-country IO (ICIO) tables are constructed by generalising the national IO tables, so as to describe the sale and purchase relationships between producers and consumers within and between economies. The most widely used ICIO tables are the Organisation for Economic Co-operation and Development (OECD) – Trade in Value Added (TiVA) and World Input-Output Database (WIOD) EORA-MRIO. WDR 2020 has used the EORA database. Needless to add, new concepts have been developed for a better understanding of value-added and GVC trade at various levels. Borin and Mancini (2019) suggest two concepts: backward and forward GVC participation. Backward GVC participation for a country implies that a country's exports embody value added it has imported from abroad. Whereas, forward GVC participation for a country implies that a country's exports are not fully absorbed in the importing country, instead are embodied in the importing country's exports to third countries. In other words, GVC participation is termed as "backward" if intermediate inputs are from the preceding stage of production; participation is "forward" if the exporter is at an early stage of production. For example, India exports aluminium tubing to Taiwan and China, where it is further used in the production of the bicycle, which is later exported. Here India's GVC participation is considered as forward, whereas China & Taiwan's GVC participation is backward. Forward and backward participation is based on the global input-output tables and strong assumptions regarding the flow of intermediate inputs (de Gortari, 2019). Though GVC phenomenon is all about firm-level international trade, backward and forward participation by countries, for now, seem to be the best tools we have to understand GVCs.

In addition, WDR 2020 has classified countries, after accounting for their sizes, into four buckets; these buckets are Commodities, Limited manufacturing, Advanced manufacturing and services, and Innovative activities. The four buckets for GVC participation are formed on the basis of (a) sectoral specialisation of exports (based on domestic value added in gross exports of primary goods, manufacturing, and business services); (b) extent of GVC participation (backward integration of the manufacturing sector i.e. backward manufacturing, measured as share of the country's total exports), where higher backward integration in manufacturing is an essential characteristic of countries entering or specialised in non-commodity GVCs; and (c) measures of innovation (intellectual property or IP receipts as a percentage of GDP, and research and development or R&D intensity, defined as its expenditure of public and private R&D as a percentage of GDP). Countries will be placed in the Commodities bucket if the manufacturing share of total domestic value added in exports is less than 60 per cent, with backward manufacturing less than 20 per cent, 10 per cent, and 7.5 per cent for small, medium, and large countries respectively. These criteria ensure that manufacturing has a small share in exports and that manufacturing has limited backward linkages. Based on the primary goods' share in total domestic value added in exports, Commodity bucket is further divided into low, limited, and high commodities if the share is less than 20 per cent, equal to or greater than 20 per cent but less than 40 per cent, and equal to or greater than 40 per cent respectively. Countries are placed in the Innovative activities bucket, if for small countries IP receipts is equal to or greater than 0.15 per cent of GDP and R&D expenditure is equal to or greater than 1.5 per cent of GDP; for medium & large countries IP receipts is equal to or greater than 0.1 per cent of GDP and R&D expenditure is equal to or greater than 1 per cent of GDP. Countries belong to the Advanced manufacturing and services bucket if the share of manufacturing and business services in total domestic value added in exports is equal to or greater than 80 per cent. Countries that remain are then placed in the bucket of Limited manufacturing. Figure 2 shows backward and forward participation of these four buckets.

**Figure 2: Backward and Forward GVC participation across taxonomy buckets**

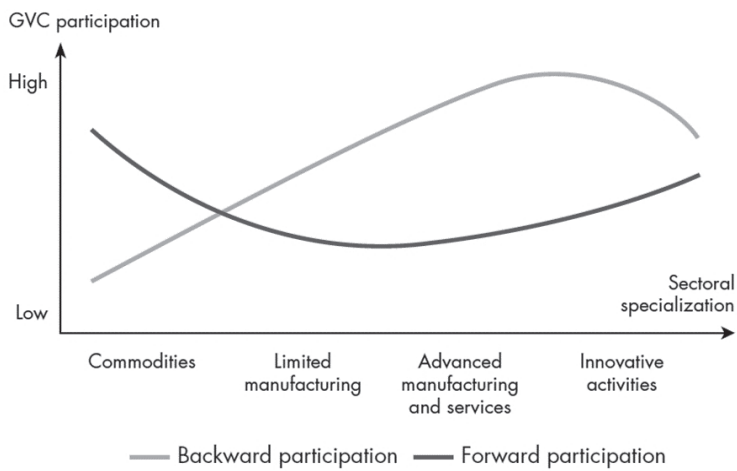

Source: WDR, 2020

It can be seen from figure 2 that on average, countries in the Commodities bucket have maximum forward participation; this is due to use of commodities in various downstream production processes. Countries placed in the bucket of Limited manufacturing generally have higher backward participation than forward participation. Highest backward participation is by countries that are placed in the Advanced manufacturing & services bucket. It is understandable that countries placed in the Innovative activities buckets have lower backward participation than Advanced manufacturing bucket, as Innovative activities are less reliant on imported inputs.

## 2.2 Trends of GVC Participation

### 2.2.1 Global Participation

Share of GVC trade in international trade, which showed steady growth in the 70s & 80s; increased quite significantly from 1990 to 2008. Although it seems to have dropped a bit in the last decade, share of GVC trade still constitutes around half of international trade.

**Figure 3: Share of GVC trade**

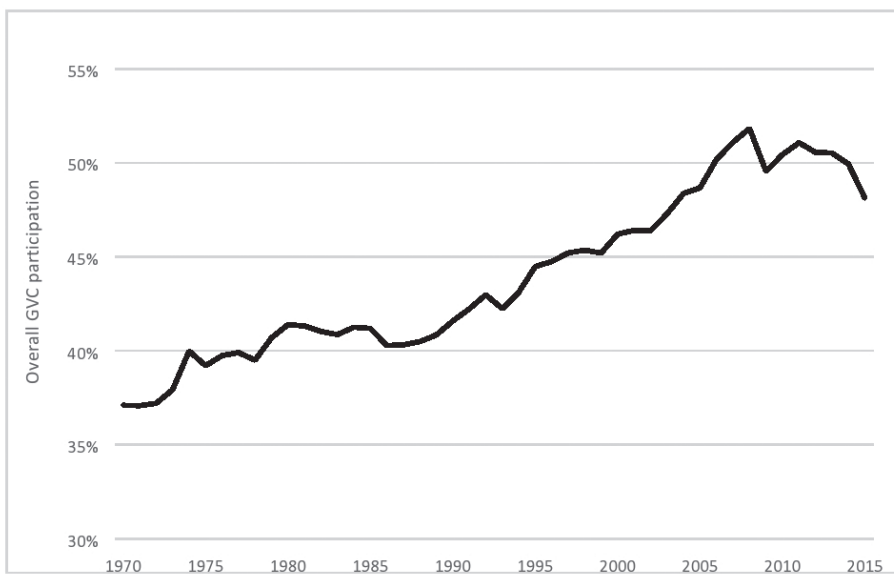

Source: WDR, 2020

WDR states that the global wave of fragmentation of production in the 1990s and 2000s was driven by a combination of factors. The information and communication technology (ICT) revolution brought forth cheaper and more reliable telecommunications, new information management software, and increasingly powerful personal computers. Manufacturing firms then found it easier to outsource and coordinate complex activities at a distance, and ensure the quality of their inputs. Another reason why firms were able to spread out production across the world was because of falling transport costs. Declining air and sea freight costs boosted the trade in goods, while services benefited from cheaper communication costs. Signing of the Generalise Agreement on Trade and Tariffs (GATT) seems to have triggered the successive rounds of trade liberalization that resulted in rapidly falling barriers to trade and investment for both developed and developing countries. Tariffs declined, especially for manufactured goods, and the gradual lowering of nontariff barriers facilitated the international trade of goods and services. Finally, the creation of the European single market - together with the integration of China, India, and the Soviet Union into the global economy - created huge new product and labour markets, and so firms could sell the same goods to more people and take advantage of economies of scale leading to the further deepening of GVCs. Supply of cheap labour encouraged profit-seeking companies to either reallocate their production facilities or find local suppliers in low-wage countries.

The global financial crisis of 2008, however, had an impact with the share of GVC trade coming down. GVC trade bounced back and grew marginally till 2011. Post-2011, however, it can be seen that GVC trade is falling. The factors behind the trade and GVC slowdown are both cyclical and structural in nature. However, in the same breath it must be mentioned that WDR (2020) notes that after the global financial crisis in 2008, the dynamics of GVC expansion have changed. Trade growth has been lower because global output growth is lower in economies such as Europe and China that account for large shares of global trade and global output. Trade has also grown at a slower pace as the trade-to-income elasticity - defined as the amount of trade generated as output rises - has decreased. This has been found to be true in large trading countries. For example, China is producing more at home, thereby becoming less reliant on imported components for its exports. The share of intermediate imports in exports of Chinese goods dropped from about 50 per cent in the 1990s to a little over 30 per cent in 2015. In the United States, a booming shale sector reduced oil imports by one-fourth between 2010 and 2015 (Constantinescu et al, 2018). Timmer et al (2016) note that though China's share in global demand was growing fast, the import intensity of Chinese final demand was well below the world average. This was because Chinese final demand increasingly not only shifted to services but also required fewer imports as more and more products were domestically produced. In the Chinese case, the share of final demand supplied by domestic industries was 88.4 per cent in 2000, dropping to a minimum of 79.9 per cent in 2006. It then increased up to 82.5 per cent in 2008 and further to 88.5 per cent in 2014. At the same time, the share of services in Chinese final demand increased steadily from 56.3 per cent in 2000, to 58.2 per cent in 2008 and further to 63.4 per cent in 2014.

*2.2.2 Indian Participation*

Understandably the Indian scenario has not been dealt with in detail in WDR (2020). Our analysis shows that during the same period in the Indian case, the share of GVC trade improved from 25 per cent to around 35 per cent, which though in line average global trend could have been much better.

If one dissects the nature of GVC participation, much of the Indian GVC trade is in the form of forward participation, which implies that we have to move up the value chain for a majority of our GVC trade. India, however, has improved its backward participation from 8 per cent to 16 per cent, which is impressive.

**Figure 4: India's GVC Trade (per cent Gross Export)**

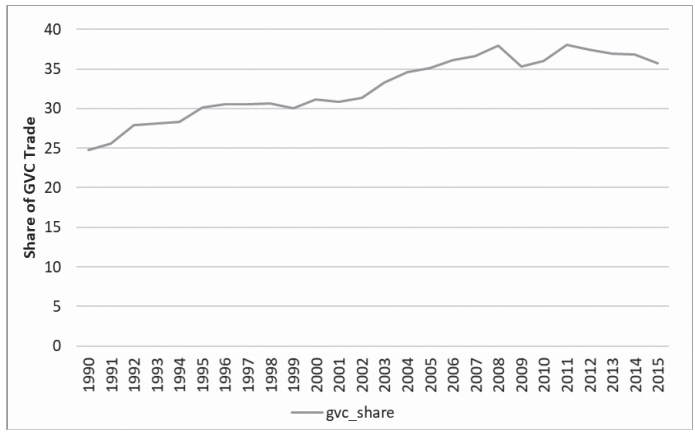

Source: EORA database

**Figure 5: India's GVC Trade (Forward vis-a-vis Backward participation)**

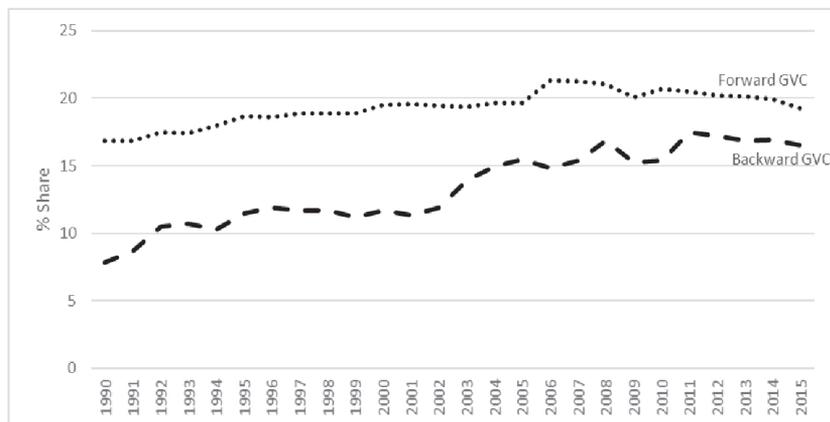

Source: EORA database

*2.2.3 Participation across Regions, Countries, Sectors, and Firms*

It can be seen from figure 6 that North American, Western European, and East Asian countries dominate the Advanced manufacturing and services bucket; whereas countries in Africa, Central Asia, and Latin America are found in the Commodities and Limited manufacturing bucket. Surprisingly, India has been put in the Advanced manufacturing & services bucket, which implies that our share of manufacturing and business services in our total domestic value added in exports is equal to or greater than 80 per cent. Given the rapid rise of business services and information technology & information technology enabled services (IT & ITES) in India, this may have been possible.

**Figure  6:  Countries in GVC buckets**

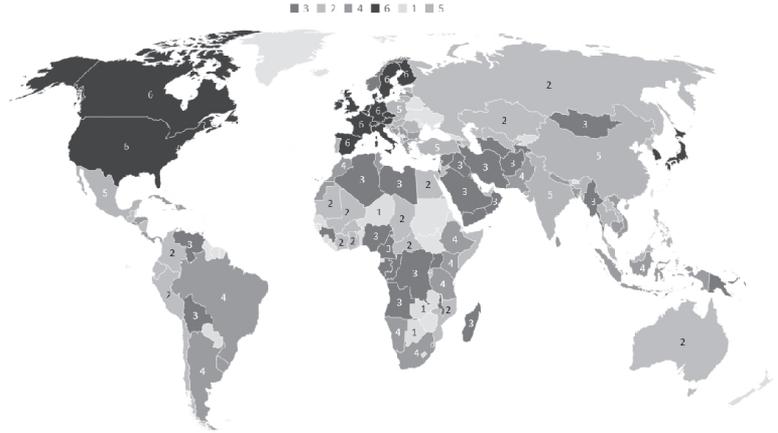

Note:   1  low participation; 2 limited commodities; 3 high commodities; 4 limited manufacturing;
         5 advanced manufacturing and services; 6 innovative activities

Source: WDR, 2020

Between 1990 and 2015, not all regions focused on global GVCs. For example, as can been from figure 7, South Asia, Latin America, Middle East, and Africa regions increased their global GVC activities; while, Europe and East Asia regions expanded their regional GVC activities.

**Figure 7: GVC pattern across regions**

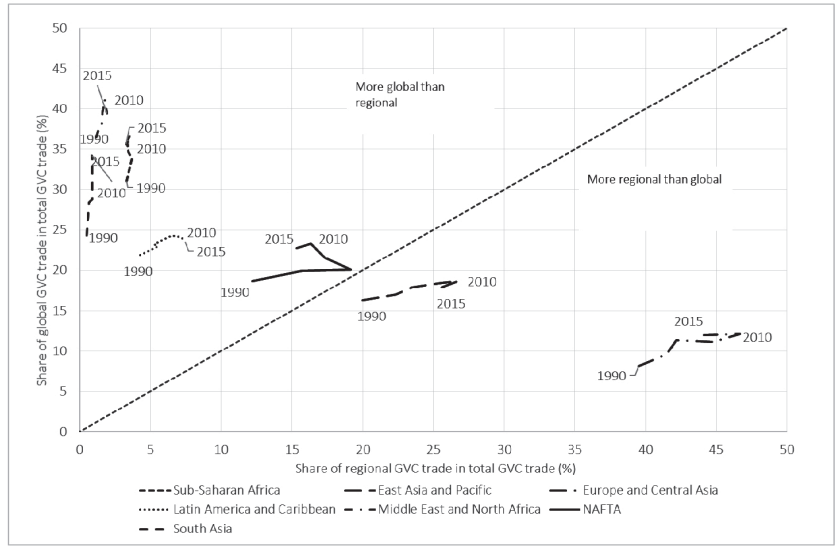

Source: WDR, 2020

GVCs expanded in developed countries, such as Germany, USA, Japan, Italy, and France through intensification of value chain in their countries; as a result of which there was growth in GVCs. On the other hand, China significantly scaled up its share of international GVC trade, which resulted in growth of GVCs. It is evident from figure 8 that during this period only a handful of countries have driven the GVC expansion.

**Figure 8: Expansion of GVCs by a handful of countries**

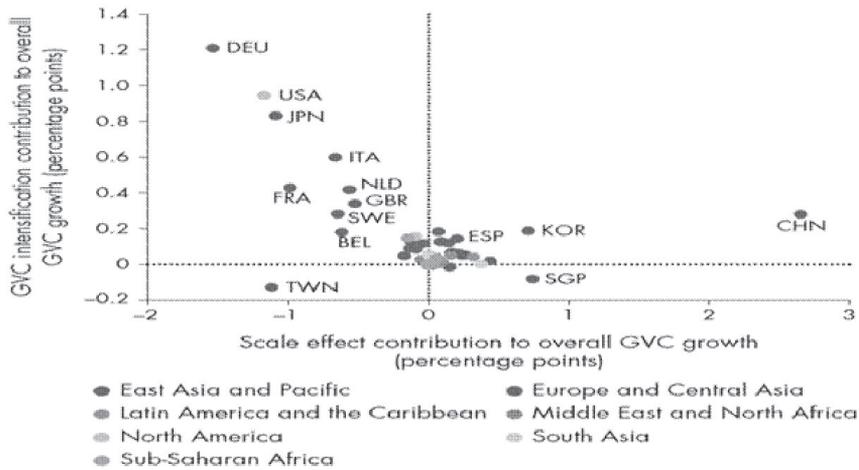

Source: WDR, 2020

**Figure 9: Sectoral GVC participation, 1995 & 2011**

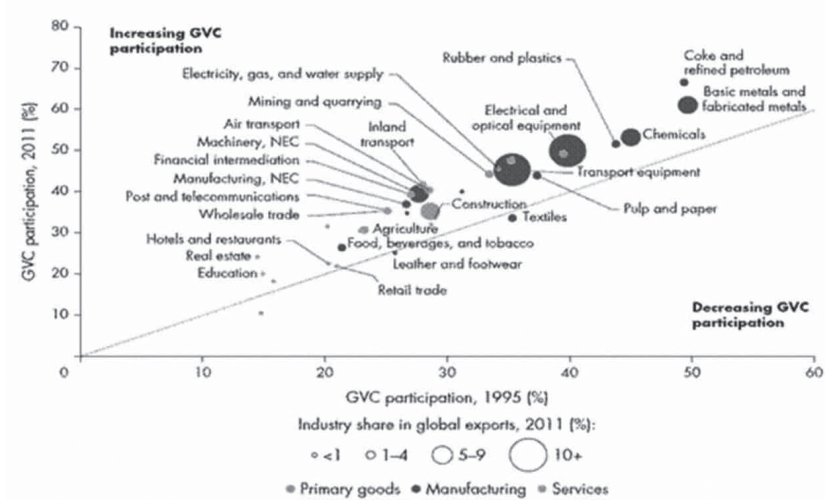

Source: WDR, 2020

    Comparatively, the sectoral composition of GVC participation is more diverse. As can be seen in figure 9, some sectors have been heavily reliant on GVCs for decades. Sectors with intensive use of resources and imported inputs such as chemicals, petroleum, metals, rubber, and plastics have increased GVC participation over time; while GVC participation in textiles and leather sectors have decreased to some extent. In services, construction and transport-related activities have also performed well.

**Figure 10: GVCs in service sector**

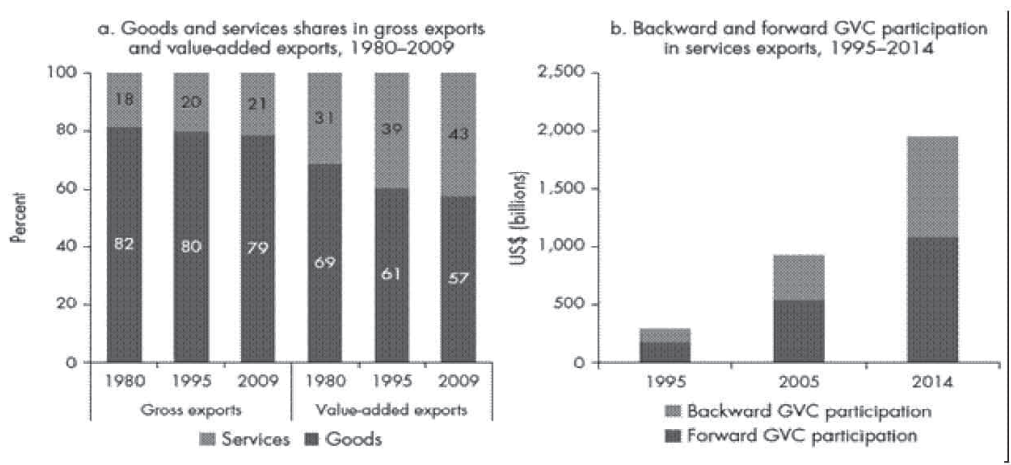

Source: WDR, 2020

During the same period, as shown in figure 10, GVCs have increased rapidly in the service sector, probably, mainly driven by multinational companies (MNCs) who are known to outsource service tasks. These services - transportation, telecommunication, and financial services – probably act as facilitation and coordination of geographically fragmented production process across sectors. Here, developed countries such as France, Germany, Italy, UK, and USA contribute to more than half of value-added exports. In this context, India's expanding role in GVC of Business services and IT&ITES is genuinely remarkable.

As is well known, firms, and not countries or industries, participate in world trade; the last two decades of economic research has seen the emergence of firm-level trade studies. In this literature, firm heterogeneity in terms of productivity leads to export & import decisions. It has been found that on average, big firms (mostly two-way traders) that constitute a small share i.e. 15 per cent of all trading firms dominate international trade (80 per cent of total trade) across countries. These firms, therefore, dominate GVC participation as "lead" firms setting up networks of upstream and downstream economic activities.

The emergence of firm-level GVCs under an environment of sunk costs, customisation, and limited contractual security leads to stickiness in the firm's behaviour especially relating to matching buyers and sellers, relationship-specific investments, flows of intangibles. In other words, the identity of the economic agents participating in a GVC is crucial, and within GVCs relationships among participants are more likely to show persistence. These relational aspects of GVCs also exemplify intra-firm trade flows. At the global level, intra-firm trade has contributed about 1/3 of world trade flows. With the firm-level approach, one can also distinguish between "producer-driven" GVCs (like

Apple) and "buyer-driven" GVCs (like Walmart) based on the complexity of products, the ability to codify transactions, and the capabilities of supply firms.

**Figure 11: GVC trade at the firm level**

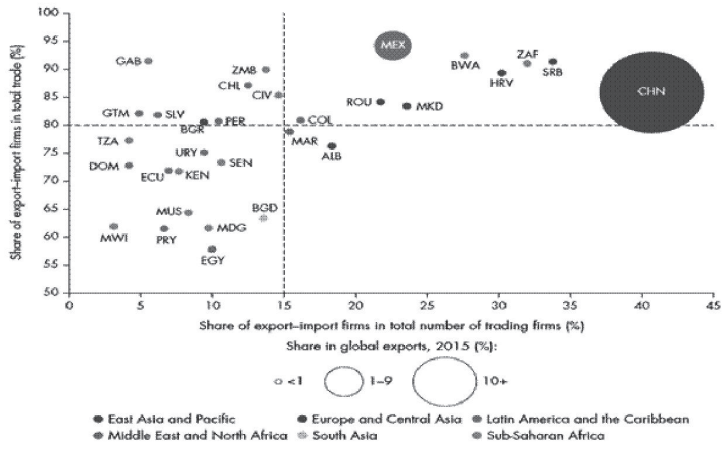

Source: WDR, 2020

Generally, MNCs use the relation-specific route either through market-seeking investment or efficiency-seeking investment, or through both of these. As figure 12 shows, there seems to be a positive correlation between foreign direct investment (FDI) inflows and GVC participation in both high-income and low & middle-income countries; however, the same thing cannot be said for FDI outflows, which are relatively high for high-income countries.

**Figure 12: Foreign Direct Investment and GVC growth**

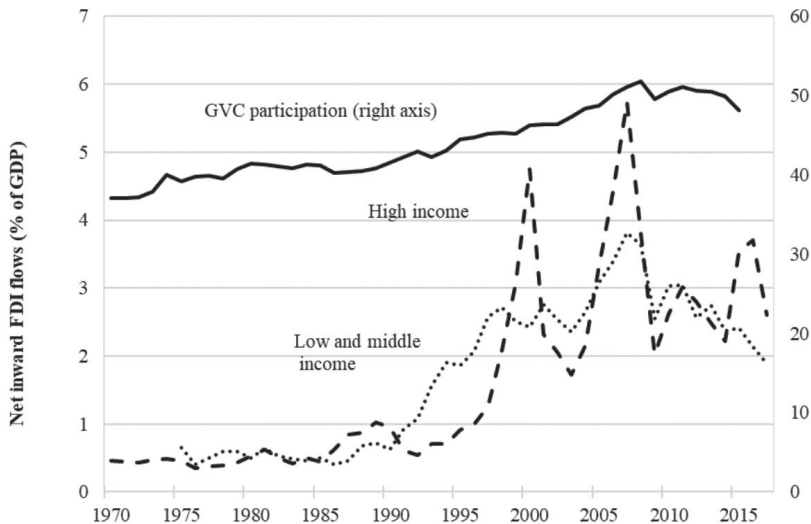

Source: WDR, 2020

From this section it is clear that GVCs have increased during the period 1990-2015, in the next section we focus on the drivers the increased this participation.

## 3. Drivers of Participation in GVCs

The rise in the global fragmentation of production processes before the financial crisis can be attributed to a combination of factors. The revolution in ICT, as shown in figure 13 not only provided cheaper and more reliable communications but also various innovative information management software, and powerful computers. This led to manufacturing and service firms outsource their various business activities and coordinate complex production activities across the globe. This increased efficiency and reduced production costs. International trade in goods was also facilitated by declining airfare and maritime freight charges; while trade in services was found to be positively affected by cheaper communications costs.

**Figure 13: ICT use and Transportation & Communication costs**

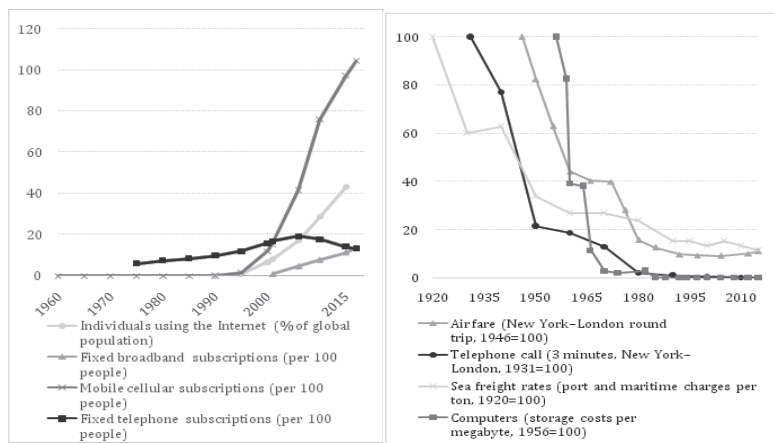

Source: WDR, 2020

During the same period, trade-related reforms undertaken by numerous countries significantly reduced tariff and non-tariff barriers, thereby enhancing international trade and GVCs activities. Creation of the European single market and integration of China into the global economy provided huge markets for new products where firms could sell products; and abundant supply of factors of production where firms benefited from economies of scale, therefore furthering the march of GVCs. Low-cost supply of abundant labour set the stage for profit-seeking firms to either reallocate their production activities or find suitable suppliers in low wage countries. There has been a decline in international trade and GVCs activities after the financial crisis which can be attributed the slower growth in international trade mainly driven by a slowdown in the global economy and particularly slower growth in large trading economies such as China, USA and EU. The slower pace and even reversal of trade reforms have adversely impacted trade growth and GVCs trading activities. It is also believed that the slowdown in GVCs activities is mainly due to the fact that the fragmentation of production in most dynamic regions and sectors has matured.

As mentioned earlier almost all countries participate to some extent in the various types of GVCs activities. Whatever be the type of participation by countries in GVCs, most of them appear to be benefiting in terms of investment flows, trade and economic growth. For example, the electronic sector in Vietnam expanded significantly in less than a decade. Vietnam, which produces 40 per cent of global mobile products of Samsung and employs 35 per cent of its global staff, has become the second-largest exporter of smartphones in the world after China. The success of Vietnam in the electronics sector can be attributed to several factors, trade liberalization policies adopted by Vietnam under the World Trade Organization (WTO) framework and various agreements with the United States (US), a favourable investment climate, and supply of young, low cost and abundant labour. These factors made Vietnam a genuine candidate for GVCs participation. As a result, international investment inflows increased including that of Samsung. The proximity of Vietnam to its regional suppliers such as China, Thailand, South Korea and Japan helped foreign investors in Vietnam to access high-quality intermediate inputs and components seamlessly and at a cheaper rate. Improved transportation and communication infrastructure during this period also helped it to engage smoothly and timely in the import and exports of goods and services.

Based on the observations of various types of participation of countries in GVCs, WDR (2020) identifies four major factors that facilitate a country's participation in the GVCs trading activities. These factors include factor endowments, market size, geographical proximity and institutional quality.

### 3.1 Factor Endowments

Traditional trade theories consider factor endowment as an important determinant of participation and specialization in GVCs. For example, countries with abundant natural resources can be associated with high forward GVCs integration, as the export of agricultural raw materials and commodities can cross several borders and be processed in a variety of downstream production processes. The success of Vietnam in the electronics sector and Bangladesh in the textile sector is mainly attributed to the supply of young, cheap and abundant labour force. Labour in Vietnam, in addition to being several years younger, can be hired at almost half the rate of their Chinese counterparts. Similarly, the average monthly wage of labour in textile sector of Bangladesh is found to be substantially lower as compared with other countries. At $197, monthly labour cost in Bangladesh is well below that of China ($270), India ($255), Vietnam ($248) and substantially lower than USA ($1864), giving an incentive to foreign multinationals to relocate their production units in Bangladesh and take advantage of low labour costs (Stitchdiary, 2017).

Apart from the availability of low wage-low skilled labour, foreign capital is also central to backward GVCs participation. Countries deficient in the capital may not participate in capital intensive

GVCs. FDI provides a solution to these countries. WDR (2020) argues that scarcity of capital prevents countries from a stronger GVCs participation in capital intensive sectors. Given the fact that most of the GVCs are monopolized by MNCs, GVCs without the involvement of MNCs is thus difficult to imagine. Vietnam's success in the electronics sector is highly dependent on investments made by Samsung. This clearly emphasises the role of FDI in boosting the domestic economy and promoting and intensifying participation in GVCs for all countries. It must be mentioned that the influence of FDI on the magnitude of backward and forward participation is not the same. Lack of FDI is said to be an important reason for the lower level of backward GVCs participation in sub-Saharan Africa (Liu & Steenbergen, 2019). FDI, however, is also associated with lower forward participation, as FDI inflows in countries may reduce the export of raw materials and thus that of intermediate services embodied in the export of resource-intensive commodities. MNCs can also promote the upstream sector in a country. They can increase the demand for domestic intermediate inputs and enable domestic suppliers to not only supply to other downstream foreign and domestic owned firms, but also to export. For instance, FDI in the textile sector of Bangladesh led to the growth of new domestic suppliers of buttons, zippers and fabrics which provided benefits to domestic firms and ensured the country's competitiveness in the global apparel sector (Kee & Tang, 2016). Such association between sectors and firms through FDI further increases the participation of countries in GVCs.

Finally, availability of a large pool of English-speaking population in a country promotes participation in service GVCs. For example, India and the Philippines are attractive offshore destinations for business and IT&ITES enabled services including complex services such as R&D.

### 3.2 Market Size

It is well known that firms in a highly protected economy, as well as those in a small market, have to rely heavily on limited domestic inputs, and face a narrow domestic demand. This restricts their trading activities as well as any possible GVC trade. Similarly, regulatory barriers on both the exports and imports in the form of tariffs or non-tariffs raise trade cost and therefore lower the volume of trade in goods and services. Trade restrictions increase the cost of imported parts and components and can reduce backward GVCs participation. High non-tariff restrictions can also increase production cost and hamper export competitiveness. For example, South Asia has high non-tariff barriers, and higher protection from imports compared to the rest of the world; no wonder, relatively, it has lower trade and participation in GVCs (Kathuria, 2018).

WDR (2020) advocates that countries must liberalize their trade policies and negotiate trade liberalization with other countries so that firms get rid not only of their dependence on limited domestic inputs but also narrow domestic demand. Over the years, trade liberalization efforts made by WTO have resulted in a significant decline in tariff and non-tariff barriers to trade. As a result, global trade and GVCs activities have expanded. Further, China's reduced tariffs on several items as a part of its accession to the WTO in 2001, promoting GVC participation for home firms but also for firms in neighbouring countries. Creation of the European single market provided firms opportunities to exploit economies of scale by selling product in this large market.

Market access, captured by tariffs, also plays a crucial role in promoting GVC participation. Those sectors facing lower tariff in destination markets show stronger backward and forward participation. WDR (2020) estimates that a one per cent reduction in an average tariff rate in destination markets increases backwards and forward GVCs integration by six and seven per cent, respectively. One way to access foreign market is through preferential access which aims at encouraging export-led growth in developing countries. Therefore, market access in destination countries such as through Everything but Arms (ETBA) initiative of European Union (EU), and African Growth and Opportunity Act (AGOA) of US can help developing countries to promote export in the short run. For example, once Bangladesh was granted duty-free and quota-free access to the European market under the ETBA initiative, Bangladesh's export of knitwear to EU doubled between 2000 and 2004 and in the same period, its export to the US increased by $30 million.

### 3.3 Geographical Proximity

The proximity between trading nations reduces trade costs and thereby production costs. Gravity model states that countries that are closer trade more than those far apart. Thus, proximity to global trading network hubs is important for participation in GVCs. For example, foreign investors in Vietnam exploited its proximity to regional suppliers of electronic inputs from China, Japan and South Korea and Singapore. Similarly, Morocco's proximity to the EU market allowed foreign investors to easily obtain automotive parts and components from suppliers in different EU countries.

Though, geographical proximity lowers trade costs and has the potential to increase participation in GVCs, inadequate physical and communications infrastructure and delays in customs clearances can ensure the opposite. A major constraint for developing countries not been able to increase their GVC participation is inefficient transport and logistics services. For example, slow and unpredictable land transport keeps most Sub-Saharan African countries out of the electronics value chain (Christ & Ferrantino, 2011). Availability of cold storage facilities enabled value chain in Ethiopian floriculture to develop (Ponte et al, 2014), while its unavailability in Bangladesh limited Bangladesh's participation in aquaculture value chains (Arvis & Marteau, 2010). Improving physical infrastructure such as roadways, airways and maritime would allow countries to enhance their participation in GVCs. In a similar manner, robust communication infrastructure allows GVCs enterprises to be seamlessly connected with their customers and suppliers.

### 3.4 Institutional Quality

Since goods and services in GVC trade cross borders multiple times, quality of institutions in each participating country form the bedrock of efficiency of the production process. Weak contract enforcement at any stage in any country can hamper the entire production processes and result in production delays. As a result GVCs may not want to enter countries that have weak institutions. Levchenko (2007) finds that exchange of intangibles, especially intellectual property rights, and relation specific GVC investments can be facilitated by improving institutional quality. Ceteris paribus, GVC participation of countries having better institutional quality has increased for those sectors that depend more on contractual enforcement (WDR, 2020). Poor institutional quality linked to land and property rights in Ghana and Cote DIvoire was found to be the major constraint to the growth of Agri-processing GVCs (Amanor, 2012).

WDR (2020) suggests that Preferential Trade Agreements (PTAs) with appropriate provisions, through financial and technical assistance, can improve domestic institutional quality and hence GVC participation. These PTAs include legal and regulatory frameworks that enforce intellectual property rights by setting rules and standardizing custom procedures. WDR (2020) gives references to PTAs that increase countries' backward GVC participation in regional trade blocs. For example, trade agreements between EU and Association of South Asian Nations (ASEAN) are positively linked to stronger backward GVCs participation of their members; whereas a weak but positive impact is seen in the case of North America Free Trade Agreement (NAFTA) (Johnson & Noguera, 2017). Lower tariffs, large FDI inflows, and stronger regulatory framework are major channels by which PTAs improve their backward GVC participation. WDR (2020) also notes that political stability is crucial for promoting backward GVCs participation.

**4. Upgrading in the Global Value Chains**

WDR (2020) states that, countries, in order to upgrade along the GVC must improve in each of the four factors important for participation in GVCs. Figure 14 shows that countries transitioning up in different categories of GVCs have increasing labour costs pre and post entry, which could imply increase in labour productivity and hence skill of the workforce.

**Figure 14: Labour costs and Capital stock – Pre and Post GVC entry**

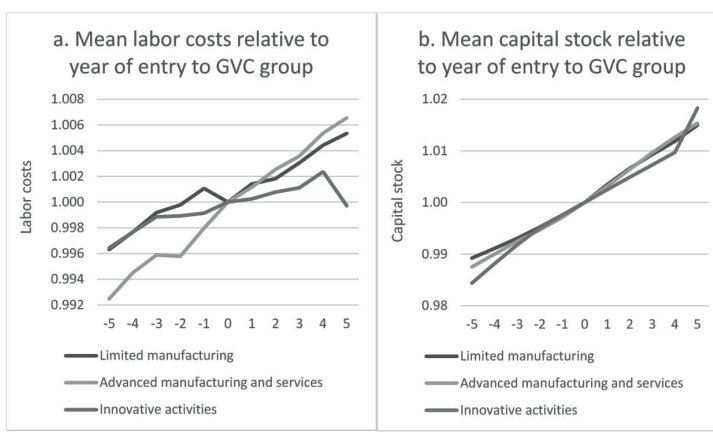

Source: WDR, 2020

Transition along the GVC is accompanied by reduction of mean manufacturing tariff as well as larger FDI inflows. This can be seen from figure 15.

**Figure 15: Manufacturing Tariff and FDI Inflows – Pre and Post GVC entry**

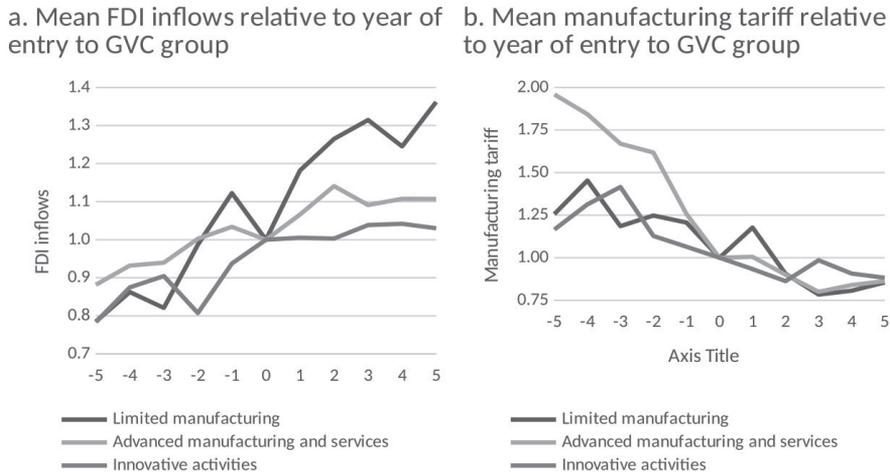

Source: WDR, 2020

Countries that have upgraded along the GVC are also seen to be enhancing transportation and logistics infrastructure, while improving their institutional quality and political stability. Transitioning of the Czech Republic, in the 1990s, from Limited manufacturing bucket to Advanced manufacturing and services bucket; and then to Innovative activities bucket after 2010, is a case in point (WDR, 2020). Various factors that have been attributed to this transitioning include its proximity to Austria and Germany, supply of low-cost skilled labour force compared to Germany – which made it an attractive destination for large FDI inflows – and, political stability particularly after the fall of Soviet Union in the 1990s. In addition, average manufacturing tariffs in Czech Republic declined to 2 per cent by 2000. In 2004, by becoming a member of the EU in 2004, it got access to a large market. Emphasis on human capital and innovation helped increase by 2007 the share of high skilled workers to 40 per cent, and R & D expenditure rose from 1.1 per cent in 2000 to 1.9 per cent of GDP in 2015. During this period, communications and digital infrastructure in the Czech Republic also expanded.

As seen from figure 16, from 1990-2015, India moved up from the Limited manufacturing bucket to the Advanced manufacturing and services bucket. Though, there has not been much mention in WDR (2020) on India's ascent; our analysis shows that India's place in the Advanced manufacturing and services bucket is mainly driven by the contribution of its services sector in the domestic value-added of export rather than the manufacturing sector. For example, Branstetter et al (2018) note that abundance of relatively low-cost software engineers in India was the reason for US MNCs to expand their R&D activity in our country. India's success in exporting software and business process outsourcing services established the quality of Indian software engineers, and US firms in these service sectors aggressively shifted some R&D activities to India. Since software increasingly became a critical input to innovation across the entire product space, manufacturers joined services firms in meeting some of their demand for software engineering skill with Indian labour. US MNCs were not investing in India to tap new technologies developed autonomously by indigenous Indian firms or to learn frontier science from pioneering Indian academic institutions. For the most part, the innovative capabilities of indigenous Indian firms are still viewed as quite limited compared to those of the MNCs in India. Neither were MNCs seeking to tap into an indigenous body of knowledge; they were, however, tapping Indian talent and integrating that talent into multinational R&D systems in which a significant amount of the intellectual leadership and direction still came from outside India[1]. However, in more mature and developed R&D centres, Indians have begun to exert intellectual leadership in some domains. Such successful integration of Indian talent into global R&D systems was possible due to a common language and the role of the Indian diaspora in the US. Mrinalini et al (2013) through their study support this finding. Their paper finds that in a sector-based classification of FDI from 2003-2009 showed that the software and IT sector had the highest inflow of FDI in R&D with a share of 50.30 per cent. As a result, more than 74 per cent share of the total jobs created in R&D in the country by FDI has been in the software and IT sector. However, patents generated in India by these MNC firms were a small fraction of their global patents. For example, the Indian R&D centres of the MNCs in the software and IT sector were granted only 749 patents compared to 129,385 patents these MNCs were granted globally. It is not clear from WDR (2020) if such R&D done by MNC affiliates in India has been accounted for while bucketing India in the Advanced manufacturing and services bucket.

**Figure 16: Transition within GVC buckets**

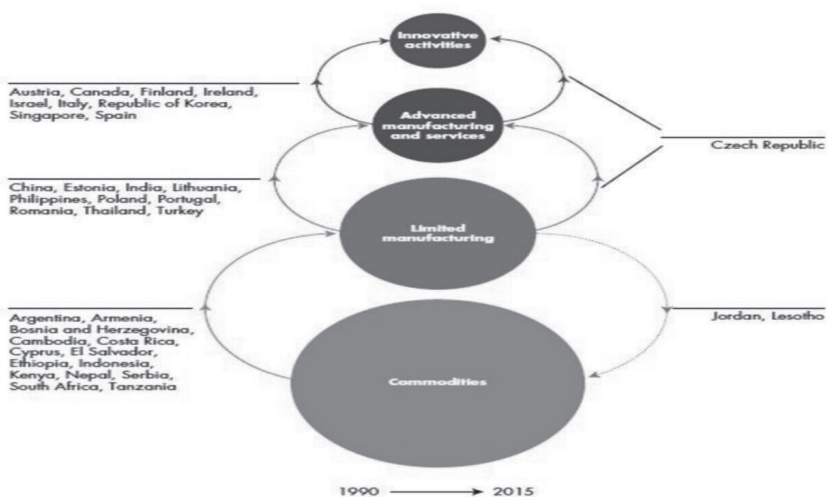

Source: WDR, 2020

---

1  This could be one of the reasons for R&D services being done in India not being captured adequately in WDR (2020).

## 5. Consequences for Development

Countries that participate in GVCs hope to reap benefits in the form of improved productivity, economic growth, increased employment, and higher wages among others. WDR (2020) highlights the importance of trade for development through the choice of the title for its report. Thus, it is equally important to understand how GVCs have impacted development. To analyse the impact of GVC participation on development, WDR (2020) has focused on the following five factors: Economic growth, Employment, Inequality, Distribution of Gains and Taxation.

### 5.1 Economic Growth

By participating into GVC, a firm can experience economic growth because of the following reasons. First, a lead firm in a GVC, generally from a developed country, outsources the task of producing different parts of the product to various firms mostly from developing countries. The arms-length contractual relationship between the firms helps the developing country firm import high-quality inputs, receive strategic suggestions to manage the production process as well as get technical knowledge for efficient production, generally from the lead firm. This can lead to higher productivity for the developing country firm while reducing the cost of production for the lead firm (Jones et al, 2005). Second, a closer and long term relationship between lead firms and the developing country firms can reduce transaction costs like searching for suppliers (for lead firms) or buyers (for their subsidiaries). WDR (2020) through figure 17 shows that suppliers who hold close ties with lead firms are 38 per cent more likely to get assistance from lead firms than others. This can lead to higher GDP per capita and growth of total export for the supplier countries.

**Figure 17: Probability of receiving assistance for firms that have close ties with lead firms**

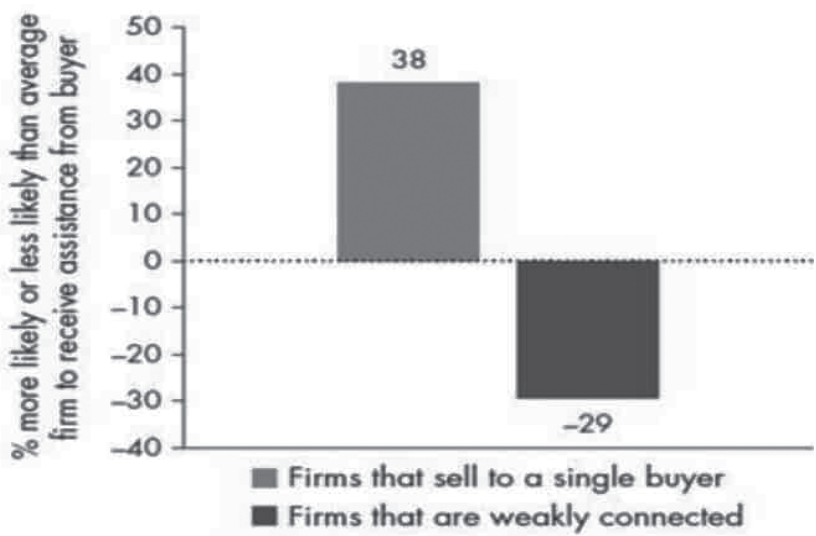

Source: WDR, 2020

Third, by engaging with lead firms and through it getting access to the global market, suppliers from developing countries can break the barrier of insufficient demand. Larger market provides the opportunity to reap the benefits of economies of scale. Fourth, a firm with lower skills can enter a GVC by contributing in Limited manufacturing. In fact, WDR (2020) shows that compared to other buckets the cumulated change in GDP per capita is highest for Limited manufacturing. This can be seen from figure 18.

**Figure 18: GDP per capita growth for GVC buckets**

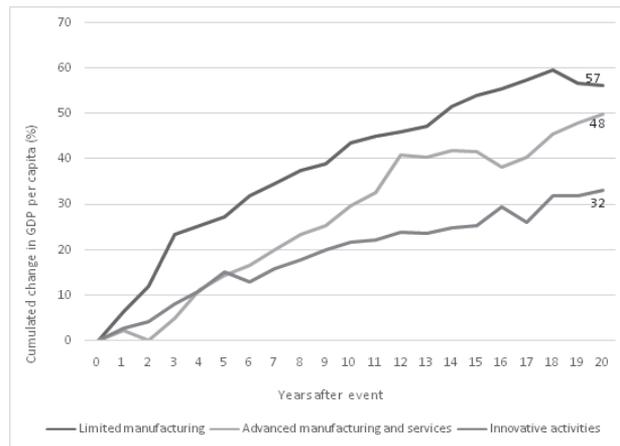

Source: WDR, 2020

WDR (2020), as shown in figure 19, highlights that backward participation has the highest increase on per capita GDP. In other words, foreign value added in the share of gross export has the most positive impact on per capita GDP growth of developing countries. This implies that countries for maximum benefit should aim for backward participation in GVCs.

**Figure 19: Impact of participation in GVC on per capita GDP**

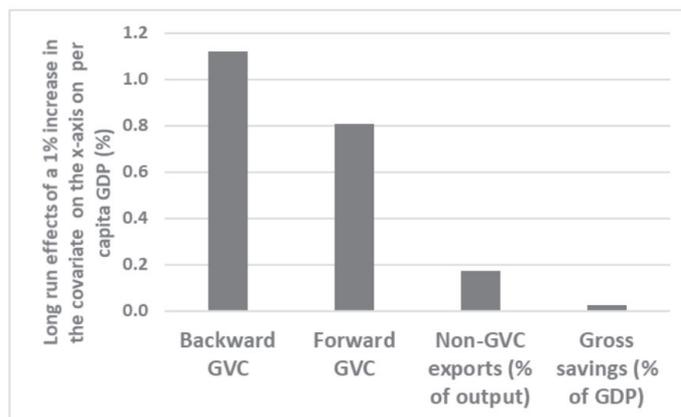

Source: WDR, 2020

## 5.2 Employment

Since productivity of GVCs are high, production activities in GVCs have been found to be more capital intensive than normal trade. This may lead to lesser employment creation as capital intensive techniques are labour saving in nature. However, since GVCs cater to the global market they have a huge scale effect, which can create employment in other sectors. In other words, GVC can reduce the worker per unit firm, but has the capability to create more firms. By shifting the focus away from local market to global market, the developing country caters to a larger market which can lead to increase in demand for labour. WDR (2020) supports this point by showing the increase in employment for Mexico, Vietnam and Ethiopia firms after they participated in GVC; for Mexico, this can be seen from figure 20.

**Figure 20: Employment pattern for firms in Mexico**

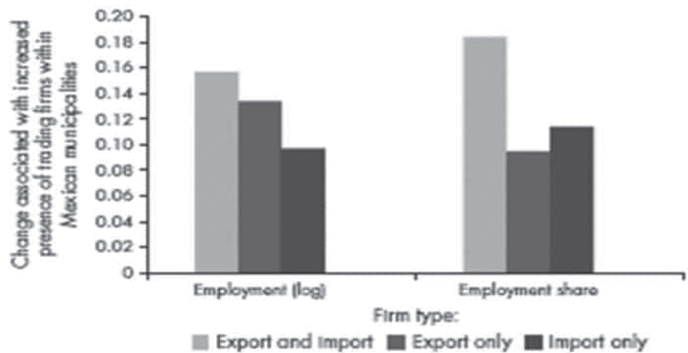

Source: WDR, 2020

WDR (2020) also highlights that the employment in these countries is rising in both the formal and the informal sectors. As one would expect, highest growth in wages are found to be when countries enter the Limited manufacturing bucket, clearly shown in figure 21.

**Figure 21: Wage growth on entering GVC buckets**

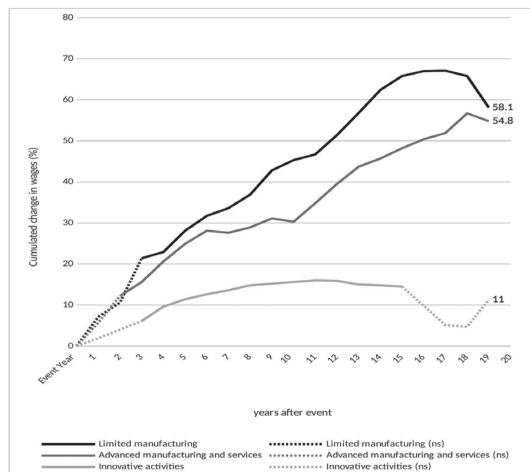

Source: WDR, 2020

Finally, GVCs are also acknowledged for providing more job opportunities for the women labour force in comparison to other trade and non-trade production activities.

**5.3 Inequality and distribution of gains**

Though on the one hand WDR (2020) paints an optimistic picture with respect to economic growth and employment creation, on the other hand, it also shows that GVCs can have a detrimental impact on both intraregional and global inequality. The first reason is the differential mark-ups charged by the developed and the developing country firms. By taking the Japanese and the Indian textile industry as representative of developed and developing countries respectively, WDR (2020) through figure 22 shows that as countries integrate into GVC, mark-ups are rising for developed countries – which more or less have backward participation; mark-ups, however, are falling for developing countries – which more often than not have forward participation.

**Figure 22: Developed and developing countries mark-up**

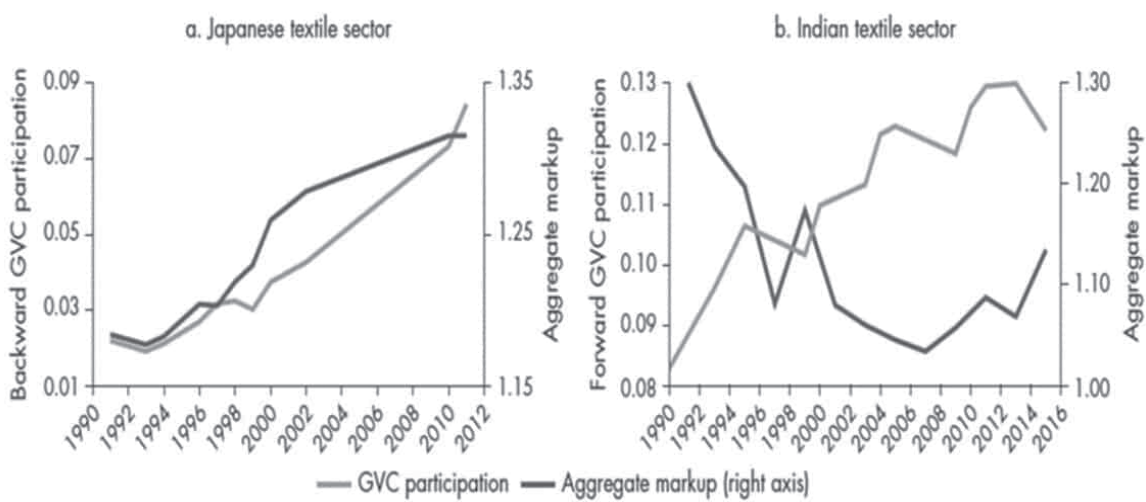

Source: WDR, 2020

It can be seen that mark-ups are not rising for all the GVC participants. So, participating in GVC is more beneficial for developed country firms compared to those in developing country firms. Second, WDR (2020) shows that labour's share of total value received from GVC activities is falling for both developed and the developing countries, which is a corollary of the labour-saving techniques used in GVCs. Thus, the distribution of the total value generated in the GVC activities is skewed towards employers of developed countries. One of the possible reasons mentioned is the skill gap between the Limited manufacturing bucket, Advanced manufacturing services bucket and Innovation activities bucket. Innovation activities and Advanced manufacturing and services are more sophisticated in nature and need specialised skills. Limited manufacturing, on the other hand, requires generic skills that can be easily learned and imitated. This increases availability of labour for Limited manufacturing, creating a downward pressure on wages. This skill gap leading to wage gap can also be a cause for regional inequality. A distinct international division of labour can be discerned with Limited manufacturing works concentrated in developing countries like Bangladesh, Vietnam, Ethiopia and Mexico; while innovation activities concentrated in developed countries like US, Germany and United Kingdom. Another essential factor causing intraregional inequality is the preference towards urban as opposed to rural areas. Cities have better infrastructure and connectivity, hence become the nucleus for industrial clusters. This creates a situation of uneven development between the urban centres and the rural peripheries. GVC activities do not appear to be solving this centre-periphery disparity. Finally, everybody participating in GVC activities does not necessarily have equal benefits. Volume of work generated in GVC activities is huge but not regular. GVC products are complex and demand high customisation, with quicker production, and are vulnerable to the sporadic market demands. As a result, employers hire contractual workers. These workers have to bear with short term contracts, lower wages and insufficient social securities. For example, Indian automobile industry, since liberalisation has created jobs in tier 1, tier 2 and tier 3 supplier firms. There is a clear distinction in the condition of the work between the tiers, with tier 1 creating formal jobs or regular employment, tier 3 creating mostly informal jobs.

### 5.4 Taxation

Another big concern that WDR (2020) raises is on the impact of GVC on taxation. Tax is the primary source of revenue for the governments and it creates the fund for public expenditures. Lead firms of GVCs, prefer tax concession to ensure higher surplus, thus taking advantage of differences between national tax systems to shift production to lower-tax jurisdictions. Countries compete by lowering corporate income tax rates and granting tax incentives such as tax holidays and preferential tax zones. In a race to the bottom, corporate income tax rates have declined by almost half since 1990. The report notes that "*Revenues from corporate income taxes are further eroded by international tax avoidance, which takes advantage of loopholes and weaknesses in the international tax architecture. In GVCs that involve affiliates of the same common corporate structure, firms can locate activities that generate high profits with relatively little input, or "substance," in jurisdictions where those profits are taxed at low rates. Such practices are legal, but they run counter to the principle of taxing activities where value is created. Firms can also manipulate transfer prices between their affiliates to shift profits to lower-tax jurisdictions.*" (WDR, 2020) The report shows that for developed as well as developing countries, corporate tax has been steadily declining over time. It also estimates that from 2012 to 2018, the revenue lost for the OECD countries is on average 0.9 per cent of the GDP and it is 1.3 per cent for the Non-OECD countries. This can be seen from figure 23.

**Figure 23: Taxation revenue losses**

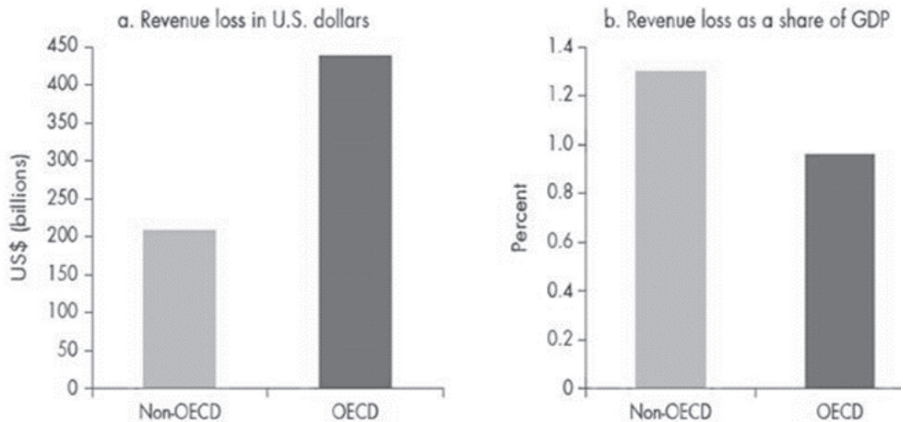

Source: WDR, 2020

## 6. Observations

WDR (2020) is sanguine about the global trade system; it notes that developing countries have benefited from the rules-based trade system, with its guarantees against trade discrimination, incentives to reform, assured market access, and dispute settlement. The international trade system is especially valuable in a GVC world. Policy action or inaction in one country can affect producers and consumers in other countries. Increasing pressure on the global trading system, manifested in protectionism and policy uncertainty, puts these benefits at risk. These pressures arise, first, from the growing symmetry in the economic size of countries and the persistent asymmetry in their levels of protection; second, from the failure to use domestic policies to address labour market dislocation and growing inequality in some advanced countries. To sustain beneficial trade openness, countries need to deepen traditional trade cooperation to address remaining barriers to trade in goods and services, as well as other measures that distort trade, such as subsidies and the activities of state-owned enterprises. Meaningful outcomes may be possible if major developing countries engage as equal partners and even leaders instead of seeking special and differential treatment. India is a major developing country; this implies that WDR (2020) calls upon India to engage as equal partners if not leaders.

The report continues that developing countries would benefit from policies that spread the jobs and earnings gains from GVC participation across society. Access to child care and training programmes support jobs for women and youth, respectively. Smallholders need assistance, such as extension services and access to finance, to integrate into agricultural value chains. GVC lead firms, labour, and governments can work together to protect workers' safety and rights. Industrial countries would benefit from adjustment policies for workers displaced by technology, trade, and the expansion of

GVCs. Placement services, training, and mobility support can help workers transition to more productive jobs. Policy can mitigate negative environmental consequences and promote the adoption of environmentally friendly technologies. Pricing the environmental costs of production and distribution appropriately will encourage conservation and cleaner technologies. In addition, regulation is needed for specific pollutants and industries. These national measures can be complemented by global cooperation on the environment and working conditions. Standardized international data will help expose poor production practices and induce firms to improve (WDR, 2020).

## 6.1 Some Concerns

It is evident that though multiple firms, from different corners of the world, are participating in GVC, the distribution of the returns and the conditions for participating are dictated by lead firms who are mostly from developed nations. GVC depicts a situation where we find many firms intensely competing with each other to capture the supplier contract, and lead firms with giant market power coordinating the entire chains. This leads to a situation where the upstream activities like producing parts and assembling becomes low value-generating activities. The unequal nature of the competition in the production process comes from not only skewed ownership of market share but also disproportionate ownership of financial capital. As a result, the cost of shifting a supplier for the lead firm becomes very low, especially when the tasks are easily replicable – which increases the number of suppliers. This gives lead firms immense power to determine the condition of the contract. For instance, in the textile industry to attract lead firms, India has to compete with Bangladesh. This competition is on the cost criteria. Eventually, the lead firm will give the contract to that firm which quotes the lowest cost for production. Figure 24 from WDR (2020) has some pointers.

**Figure 24: Employment generated by GVC vis-a-vis Export**

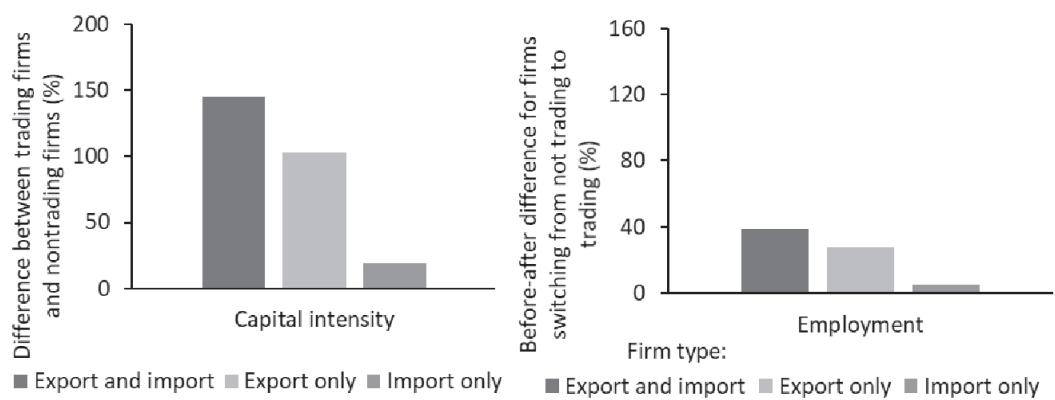

Source: WDR, 2020

It is clear from figure 24 that the ratio of employment generated to capital intensity may not be highest for GVC firms. From the figure, it seems that this ratio is the highest for export firms. This implies that, ceterus paribus, if a country has the option to export without joining a GVC then that path may be more employment generating.

Taxation of GVC firms is another thorny issue. MNCs register their company under different sectors depending on which sector the tax is lowest (Amazon registers in many places as e-commerce company, some places as logistic company and some places as retail company). Often, they earn in one country and file their returns as a different establishment in another country which has lower tax rates. It is not clear if for developing countries there is a way out of this issue.

WDR (2020) highlights the fact that GVCs are recruiting more women workers. A possible explanation for this can be that GVCs are creating more informal work than formal work particularly in developing countries like Vietnam, Bangladesh, and India (Mezzadri, and Fan 2018). These self-employed women workers participate in the GVC through subcontracting. It is, however, unclear, whether participating in GVCs gives them better working conditions, higher wages, better social security and secured job contracts. The example of the Barber tribes of Morocco, producing argan oils shows that GVCs by themselves do not ensure better working conditions (Meagher,2019).

## 7. Way Forward

WDR (2020) prescribes that GVCs still offer developing countries a clear path to progress and that developing countries can achieve better outcomes by pursuing market-oriented reforms specific to their stage of development. The report mentions that participation in GVCs can deliver a double dividend. First, firms are more likely to specialize in the tasks in which they are most productive. Second, firms are able to gain from connections with foreign firms, which pass on the best managerial and technological practices. As a result, countries enjoy faster income growth and falling poverty. Implicit in their prescription is the assumption that the process of trade reforms at each country level as well as the process of globalization will not reverse. Recent international developments clearly question this assumption. We seem to be headed into a world with few trading blocks and multiple bilateral free trade agreements among countries. Thus it is an open question if gains obtained by participation in GVCs from 1990-2015 can be replicated in future. In addition, as we have argued above more empirical proof is required to show that lead firms in GVCs do not enjoy disproportionate bargaining power than their suppliers. Also, as the report itself shows, the ratio of employment generated to capital intensity may not be highest for GVC firms. The report also does not provide solutions to the vexed issue of taxation of GVCs. These are some of the questions that need to be answered before we can accept the WDR (2020) prescriptions.

On the contrary, if one accepts the WDR prescription that GVCs offer India a clear path to progress, without question, then there are certain India specific constraints that need to be addressed before India can reap the benefit of GVCs. The report mentions that factor endowments, market size, geographical proximity and institutional quality are four main drivers for participation. Being the second most populated country in the world, India is well endowed with labour as well as a huge market. However, though India is closer to GVC hubs such as China and Vietnam, port and road bottlenecks continue, which is an issue that needs urgent attention. Finally, on the quality of institutions, Subramanian (2007) notes that around the world as countries grow, political and economic institutions tend to improve. As people become richer, they demand more from their public institutions—better public services, more security and law and order, and greater political participation. However, in the Indian case, he finds there is no evidence of improvements in the average quality of institutions over time; if anything, the evidence leans in the other direction. This clearly shows that India may have to get its house in order before it expects further benefits from GVCs.

*This commentary is the outcome of a panel discussion held at CDS on January 17, 2020. We are grateful to Prof. Sunil Mani, Prof. Sudip Chaudhuri and other participants at the panel discussion for their comments which has benefited this commentary. We are solely responsible for any errors that remain.*

***Chidambaran G. Iyer*** *is Associate Professor at the Centre for Development Studies, Trivandrum. His areas of specialization include, Technology, Innovation, Productivity, Spillovers.*

*Email : cgiyer@cds.ac.in*

***Rajkumar Byahut****,* ***Sourish Dutta*** *and* ***Manikantha Nataraj*** *are Doctoral Scholars at the Centre for Development Studies, Trivandrum.*

**PREVIOUS PAPERS IN THE SERIES**

1. **Dimensions of India's Economy:** *As seen through the Economic Survey 2017-18 and the Union Budget 2018-19.*
   Manmohan Agarwal, Sunandan Ghosh, M Parameswaran, Ritika Jain, P Seenath, Hrushikesh Mallick, Vinoj Abraham, Udaya S Mishra, Sunil Mani, P L Beena. March 2018.

2. **Dimensions of India's Intellectual Property Right System. How Many Patents are Commercially Exploited in India?**
   Sunil Mani. May 2018.

3. **An Uncertain Shift from 'Protectionism' to 'Empowerment'? Probing the Decision by NORKA to Recruit Women Domestic Workers for Kuwait**
   Praveena Kodoth, November 2018.

4. **Protectionism: US Tariff Policy and India's Response**
   **Part I:** Manmohan Agarwal, **Part II:** Sunandan Ghosh, December 2018.

5. **Industrial Investment Intention and Implementation in India: Broad Trends and Patterns at the State-level**
   Ritika Jain, March 2019.

6. **Monetary Policy Journey to Inflation Targeting**
   Manmohan Agarwal & Irfan Ahmed Shah, April 2019.

7. **Gender-Based Cyber Violence against Women in Kerala: Insights from Recent Research**
   J. Devika, May 2019.

8. **SWACHH BHARAT - 2019**: **Will Rural India be ODF/SWACHH?**
   G. Murugan, May 2019.

9. **Dimensions of Indian Economy** *As seen through the Economic Survey 2018-19 and the Union Budget 2019-20*
   K. P. Kannan, P.L. Beena, M. Parameswaran, Vinoj Abraham, G. Murugan, Udaya Shankar Mishra, Hrushikesh Mallick, Sunil Mani**,** August 2019.

10. **Are Medicine Prices High and Unaffordable after TRIPS? Evidence from Pharmaceutical Industry in India**
    Sudip Chaudhuri, December 2019.

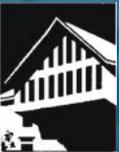

**Centre for Development Studies**
(Under the aegis of Govt. of Kerala & Indian Council of Social Science Research)
Prasanth Nagar, Ulloor
Thiruvananthapuram 695 011
Kerala, India

Phone : 0471 - 2774200, 2448881, 2442481, 2448412
Fax: 0471 - 2447137
Website: www.cds.edu